\documentclass[twocolumn,showpacs,prl]{revtex4-1}
\usepackage{graphicx}
\usepackage{amsmath}
\usepackage{amssymb}

\begin{document}

\title{Scheme to Achieve Silicon Topological Photonics}
\author{Long-Hua Wu}
\email{WU.Longhua@nims.go.jp}
\affiliation{$^1$International Center for Materials Nanoarchitectonics (WPI-MANA),
National Institute for Materials Science, Tsukuba 305-0044, Japan \\
$^2$Graduate School of Pure and Applied Sciences, University of Tsukuba, Tsukuba 305-8571, Japan}
\author{Xiao Hu}
\email{HU.Xiao@nims.go.jp}
\affiliation{$^1$International Center for Materials Nanoarchitectonics (WPI-MANA),
National Institute for Materials Science, Tsukuba 305-0044, Japan \\
$^2$Graduate School of Pure and Applied Sciences, University of Tsukuba, Tsukuba 305-8571, Japan}

\date{\today}

\begin{abstract}
We derive in the present work
topological photonic states purely based on silicon, a conventional dielectric
material, by deforming a honeycomb lattice of silicon cylinders into a
triangular lattice of cylinder hexagons. The photonic topology is associated
with a pseudo time reversal (TR) symmetry constituted by the TR
symmetry respected in general by the Maxwell equations and the $C_6$ crystal
symmetry upon design, which renders the Kramers doubling in the present
photonic system with the role of pseudo spin played by the circular
polarization of magnetic field in the transverse magnetic mode. We solve
Maxwell equations, and demonstrate new photonic topology by revealing
pseudo spin-resolved Berry curvatures of photonic bands and helical edge
states characterized by Poynting vectors.
\end{abstract}

\maketitle

\emph{Introduction.---}The discovery of quantum Hall effect (QHE) opened a new
chapter of condensed matter physics with topology as the central concept
\cite{QHE,TKNN,KaneRMP,SCZhangRMP,QNiuRMP,Haldane1988,Kane2005,Bernevig2006,Hsieh2008,
RYuSci2010,QKXueSci2013}. Topological states are not only interesting in an
academic point of view, but also expected to yield significant impacts to
applications because robust surface (or edge) states protected by bulk
topology provide new possibilities for spintronics and quantum computation
\cite{Liang2013,MacDonald2012,CNayakRMP,TewariJPCM2013,Beenakker2013,WuSTAM2014}.
However, topological matters confirmed so far are still limited in number, and
most of them exhibit topological properties only at very low temperatures,
which hinders their better understanding and manipulation indispensable for
practical applications.

Photonic crystals are optical analogues of solids with lattice of atoms
replaced by medium of periodic electric permittivity and/or magnetic
permeability \cite{EYab1987}.  Metamaterials are designed to generate
electromagnetic (EM) properties such as negative index, magnetic lens, and so
on, which are not available in nature \cite{Pendry99}.  Recently it has been
recognized that topological states characterized by unique edge
propagations of EM wave can be realized in photonic crystals based on
gyromagnetic materials under external magnetic field, bi-anisoctroic
metamaterials with coupled electric and magnetic fields where bi-anisotropy
acts as effective spin-orbit coupling, and coupled resonator optical
waveguides (CROWs)
\cite{Haldane2008,SoljacicPRL08,SoljacicNature09,FangNPhoton12,ShvetsNM12,Shvets14,ChongPRL13,RechtsmanNature13,YFChen14,HafeziNPhys11,HafeziNPhoton13,Ochiai2012}.
(for a review see \cite{SoljacicNPhoton14}).

\begin{figure}[t]
  \centering
  \includegraphics[width=\linewidth]{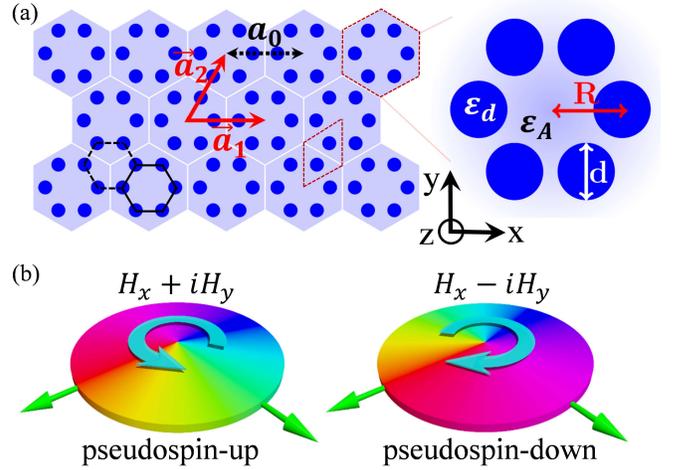}
  \caption{(a) Schematics of a triangular photonic crystal of "artificial
  atoms" composed by six cylinders of dielectric material with dielectric constant $\epsilon_d$ along the $z$
  direction embedded in the environment of dielectric constant $\epsilon_A$. Red dashed rhombus and hexagon are primitive cells of honeycomb
  and triangular lattices respectively. Solid black hexagon labels an
  artificial atom, while dashed black one marks interstitial regions among
  atoms. $\vec{a}_1$ and $\vec{a}_2$ are lattice vectors, and $a_0$ is the lattice
  constant of triangular lattice. Right panel: enlarged view of hexagonal
  cluster with $R$ the length of hexagon edge and $d$ the diameter of
  cylinders. (b) Pseudo spin states of the present photonic system associated
  with the left- and right-hand circular polarizations of magnetic fields
  $H_x\pm iH_y$ of the TM mode.}
  \label{photocrystal}
\end{figure}

In the present work, we propose a two-dimensional (2D) photonic crystal purely
made of silicon, a conventional dielectric material.  We notice that a
honeycomb lattice is equivalent to a triangular lattice of hexagonal clusters
composed by six neighboring sites, and that, taking this larger hexagonal unit
cell instead of the primitive rhombic unit cell of two sites (see
Fig.~\ref{photocrystal}(a)), the Dirac cones at K and K' points in the first
Brillouin zone of honeycomb lattice are folded to doubly degenerate Dirac
cones at $\Gamma$ point. It is then intriguing to observe that at the $\Gamma$
point there are two 2D irreducible representations in $C_6$ symmetry group
associated with odd and even parity respective to spatial inversion operation.
Based on these properties, we propose to open a topologically nontrivial
band gap by deforming the honeycomb lattice in the way keeping hexagonal
clusters and preserving the $C_6$ rotation symmetry (see
Fig.~\ref{photocrystal}(a)).  Explicitly, we reveal by solving the Maxwell
equations that harmonic EM modes hosted by the hexagonal cluster,
working as "artificial atom" in the present scheme, exhibit electronic orbital-like $s$-, $p$-, and
$d$-wave shapes and form photonic bands. We clarify that there is a pseudo
time reversal (TR) symmetry constituted by the TR symmetry
respected by the Maxwell equations and the $C_6$ crystal symmetry upon design,
which behaves in the same way as TR symmetry in electronic systems and renders
the Kramers doubling in the present photonic system. This intimately gives the
one-to-one correspondence between the left- and right-hand circular
polarizations of magnetic field in the transverse magnetic (TM) mode and the
up- and down-spin states of electrons. Evaluating the Berry curvatures of
photonic bands and the edge states for finite systems, we demonstrate the
emergence of topological phases. With the simple design free of requirement
on any external field and gyromagnetic or bi-anisotropic materials, the
present topological photonic states purely based on silicon are expected very
promising for future applications.

\emph{Artificial atom and pseudo spin.---}Let us consider harmonic TM modes of
EM wave, namely those of finite in-plane $H_x$, $H_y$, out-of-plane $E_z$
components and with others being zero, in a dielectric medium (for coordinates
see Fig.~\ref{photocrystal}(a)).  For simplicity, the real electric permittivity
is taken frequency independent in the regime under consideration. The master
equation for a harmonic mode of frequency $\omega$ is then derived from the
Maxwell equations \cite{PhotonicBook}
\begin{equation}
  \left[\frac{1}{\varepsilon(\mathbf{r})}\nabla\times\nabla\times\right]E_z(\mathbf{r}) =
  \frac{\omega^2}{c^2}E_z(\mathbf{r}),
  \label{eq:master}
\end{equation}
with $\varepsilon(\mathbf{r})$ the position-dependent permittivity and $c$ the
speed of light.  The transverse components of magnetic field are given by the
Faraday relation ${\bf H}=-i/(\mu_0\omega)\triangledown\times {\bf E}$, where
the magnetic permeability $\mu_0$ is presumed as that of vacuum. The Bloch
theorem applies for the present
system when $\varepsilon(\mathbf{r})$ is periodic in space as shown
schematically in Fig.~\ref{photocrystal}(a). It is reminded however that the
master equation (\ref{eq:master}) describes the EM waves instead of electrons
carrying on the spin degree of freedom, with the most prominent difference
lying at the response upon TR operation.

\begin{figure}[t]
  \centering
  \includegraphics[width=\linewidth]{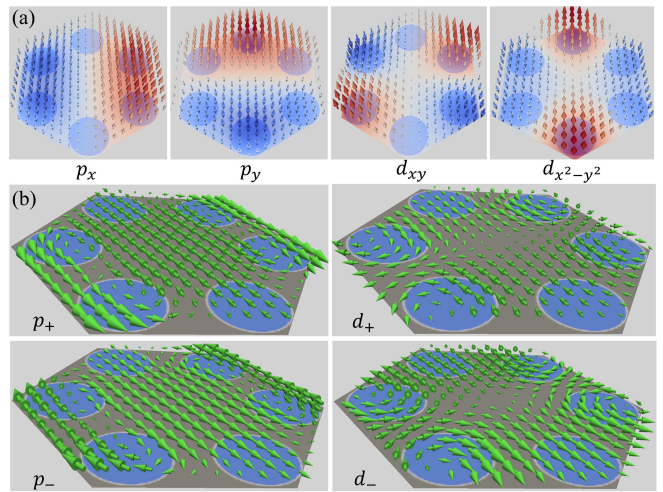}
  \caption{(a) Electric field
  $E_z$ of the $p_x/p_y$ and $d_{xy}/d_{x^2-y^2}$ photonic orbitals hosted
  by the artificial atom. (b) Two pseudo spin states corresponding to
  the left- and right-hand circular polarizations of magnetic fields $H_x\pm
  iH_y$ associated with $E_z$ fields of positive and negative angular
  momenta: $p_{\pm}$ and $d_{\pm}$.}
  \label{pseudospin}
\end{figure}

Our photonic crystal is made of cylinders of dielectric material with
relative permittivity $\varepsilon_d$ (such as silicon) parallel to the $z$
axis embedded in a dielectric background of $\varepsilon_A$ (such as air) as
schematically shown in Fig.~\ref{photocrystal}(a). Six dielectric
cylinders of diameter $d$ form a hexagon with edge length $R$. Our
discussions below are for states uniform along the $z$ axis, which reduces the
problem to 2D.  We start from a honeycomb lattice of dielectric cylinders, and
deform it in the way keeping hexagonal clusters composed by six neighboring
cylinders and the $C_6$ symmetry. Now the alignment of dielectric cylinders is
more convenient to be considered a triangular lattice of hexagonal "artificial
atoms". There are two 2D irreducible representations in the $C_6$ symmetry
group associated with the triangular lattice: $E'$ and $E''$ with basis
functions $x/y$ and $xy/(x^2-y^2)$, corresponding to odd and even spatial
parities respectively \cite{Dresselhaus2008}. As can be seen from $E_z$ fields
in Fig.~\ref{pseudospin}(a) obtained by solving Eq.~(\ref{eq:master}), the
artificial atoms carry $p_x/p_y$ and $d_{xy}/d_{x^2-y^2}$ orbitals, with the
same symmetry as those of electronic orbitals of conventional atoms in solids.

We now examine matrix representations of $\pi/3$ rotation
and its combinations for the basis functions of $p_x/p_y$ and
$d_{xy}/d_{x^2-y^2}$. Since $p_x/p_y$ behave in the same way as $x/y$, it is
easy to see
\begin{equation}
  D_{E'}(C_6)
  \left(\begin{array}{c}
    p_x\\p_y
  \end{array}\right)=
  \left(\begin{array}{cc}
    \frac{1}{2} & -\frac{\sqrt{3}}{2} \\
    \frac{\sqrt{3}}{2} & \frac{1}{2}
  \end{array}\right)
  \left(\begin{array}{c}
    p_x\\p_y
  \end{array}\right).
\end{equation}
It is noticed that
$\mathcal{U}=[D_{E'}(C_6)+D_{E'}(C_6^2)]/\sqrt{3}=-i\sigma_y$ with
$D_{E'}(C_6^2) \equiv D^2_{E'}(C_6)$ is associated with the $\pi/2$ rotation
of $p_x/p_y$ ($\sigma_y$ being the Pauli matrix). Therefore,
$\mathcal{U}^2(p_x,p_y)^{\rm T}=-(p_x,p_y)^{\rm T}$, which is consistent with
the odd parity of $p_x/p_y$ with respect to spatial inversion. Similarly, one
has
\begin{equation}
  D_{E''}(C_6)
  \left(\begin{array}{c}
    d_{x^2-y^2}\\d_{xy}
  \end{array}\right)=
  \left(\begin{array}{cc}
    -\frac{1}{2}&-\frac{\sqrt{3}}{2}\\
    \frac{\sqrt{3}}{2}&-\frac{1}{2}
  \end{array}\right)
  \left(\begin{array}{c}
    d_{x^2-y^2}\\d_{xy}
  \end{array}\right),
\end{equation}
which, in the matrix form, is same as $D_{E'}(C_6^2)$ because the
basis functions are now bilinear of $x/y$. It is then straightforward to check
that $[D_{E''}(C_6)-D_{E''}(C_6^2)]/\sqrt{3} =\mathcal{U}$ is associated with
a $\pi/4$ rotation of $d_{xy}/d_{x^2-y^2}$, which yields
$\mathcal{U}^2(d_{x^2-y^2},d_{xy})^{\rm T}=-(d_{x^2-y^2},d_{xy})^{\rm T}$.

We compose the antiunitary operator $\mathcal{T} = \mathcal{UK}$ where
$\mathcal{K}$ is the complex conjugate operator associated with the TR
operation respected by Maxwell systems in general. Since $\mathcal{T}^2 = -1$ is guaranteed
by $\mathcal{U}^2 = -1$,  $\mathcal{T}$ can be taken as a pseudo TR operator
which provides Kramers doubling in the same way of electronic systems. It is
clear that the crystal symmetry plays an important role in this pseudo TR
symmetry \cite{Fu2011}.

The two pseudo spin states are given by
\begin{equation}
p_{\pm}=(p_x \pm i p_y)/\sqrt{2}; \hspace{5mm} d_{\pm}=(d_{x^2-y^2} \pm id_{xy})/\sqrt{2},
\label{eq:basis}
\end{equation}
which are related to the above basis functions by unitary transformation.
Namely the pseudo spin up and down correspond to the wave functions of $E_z$
fields of positive and negative angular momenta, or equivalently the left- and
right-hand circular polarizations of in-plane magnetic fields defined by $H_x
\pm i H_y$ as shown in Figs.~\ref{photocrystal}(b) and \ref{pseudospin}(b).

\begin{figure}[t]
  \centering
  \includegraphics[width=\linewidth]{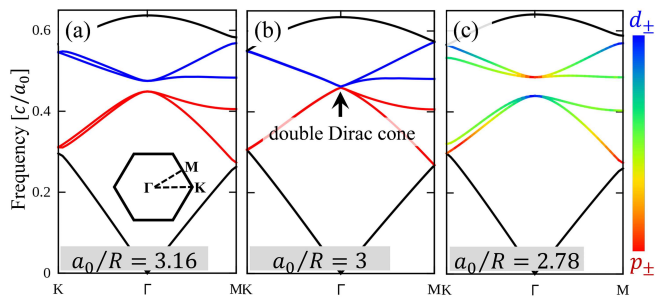}
  \caption{Dispersion relations of TM mode for the 2D photonic crystals
  with $\varepsilon_d = 11.7$, $\varepsilon_A = 1$ and $d = 2R/3$ for (a)
  $a_0/R=3.16$ (Inset: Brillouin zone of triangular lattice), (b) $a_0/R=3$
  and (c) $a_0/R=2.78$. Blue and red are for $d_\pm$ and $p_\pm$ bands
  respectively, and rainbow for hybridization between them. The case of $a_0/R=3$ corresponds exactly to
  the honeycomb lattice of individual cylinders. Diameter of individual
  cylinders is kept the same for all calculations since its variation
  influences photonic bands slightly in quantitative manners.}
  \label{pdispersion}
\end{figure}

Pseudo spins discussed so far in photonic systems include bonding/antibonding
states of electric and magnetic fields \cite{ShvetsNM12,Shvets14}, left-hand/right-hand
circular polarizations of EM waves \cite{YFChen14}, and
clockwise/anticlockwise circulations of light in CROWs
\cite{HafeziNPhys11,HafeziNPhoton13}.
The first two implementations involve sophisticated metamaterial structures and thus
require delicate fabrications. While the last one only uses silicon fibers,
and thus easy to prepare, a break in one of coupling loops at perimeter
induces a "magnetic impurity" which destroys the topological helical edge
states.

\emph{Photonic bands.---}Now we calculate the photonic band
dispersions described by the master equation (\ref{eq:master}) imposing
periodic boundary conditions along unit vectors $\vec{a}_1$ and $\vec{a}_2$
given in Fig.~\ref{photocrystal}(a).  As shown in Fig.~\ref{pdispersion}, double
degeneracy in the band dispersions appears at $\Gamma$ point, which can be
identified as $p_{\pm}$ and $d_{\pm}$ states, consistent with the symmetry
consideration.  For large lattice constant $a_0$, the photonic band below
(above) the gap is occupied by $p_{\pm}$ ($d_{\pm}$) states (see
Fig.~\ref{pdispersion}(a) for $a_0/R = 3.16$).

Reducing the lattice constant to $a_0/R = 3$, the $p$ and $d$ states become
degenerate at $\Gamma$ point, and two Dirac cones appear as shown in
Fig.~\ref{pdispersion}(b). This is because that at this lattice constant the
system is equivalent to honeycomb lattice of individual cylinders, and the
doubly degenerate Dirac cones are nothing but those at K and K' point in the
Brillouin zone of honeycomb lattice based on the primitive rhombic unit cell
of two sites \cite{Ochiai2012}.

When the lattice constant is further reduced, a global photonic band gap is
reopened as shown in Fig.~\ref{pdispersion}(c) for $a_0/R=2.78$. Now the $E_z$
field at low-(high-)frequency side of the band gap exhibits $d_\pm$($p_\pm$)
characters around the $\Gamma$ point, opposite to the order away from the
$\Gamma$ point. Namely a band inversion takes place upon reducing the lattice
constant in the present photonic lattice. Quantitatively, the band gap is
$\Delta\omega=13.8$ THz at $\omega=138.75$ THz with $a_0=1$ $\mu$m, with all
the quantities scaling with the lattice constant.

\begin{figure}[t]
  \centering
  \includegraphics[width=\linewidth]{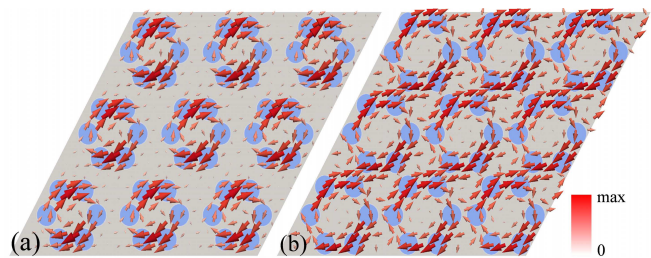}
  \caption{Real-space distributions
  of the time-averaged Poynting vector associated with the pseudo spin-down state at $\Gamma$
  point below the photonic gap: (a) $a_0/R=3.16$ in the trivial regime and (b) $a_0/R=2.78$
  in the topological regime.}
  \label{poynting}
\end{figure}

In order to see what happens in the system around the band inversion, we check
the real-space distribution of Poynting vector $\vec{S} =
\text{Re}[\vec{E}\times\vec{H}^*]/2$ averaged over a period
$\tau=2\pi/\omega$, which describes the energy flow in the present EM system.
It is found that the Poynting vector is circling around individual atoms as
shown in Fig.~\ref{poynting}(a) for $a_0/R = 3.16$, with the chirality of
Poynting vector corresponding to the pseudo spin (Poynting vector with pseudo
spin-up is not shown explicitly). The EM energy flows around individual
atoms, characterizing a conventional "insulating" state.  At $a_0/R=2.78$,
namely after the band inversion, the Poynting vectors are much enhanced in
interstitial regimes as shown in Fig.~\ref{poynting}(b).  It is in a sharp
contrast to the case in Fig.~\ref{poynting}(a), and implies an unconventional
insulating state.

\begin{figure}[t]
  \centering
  \includegraphics[width=\linewidth]{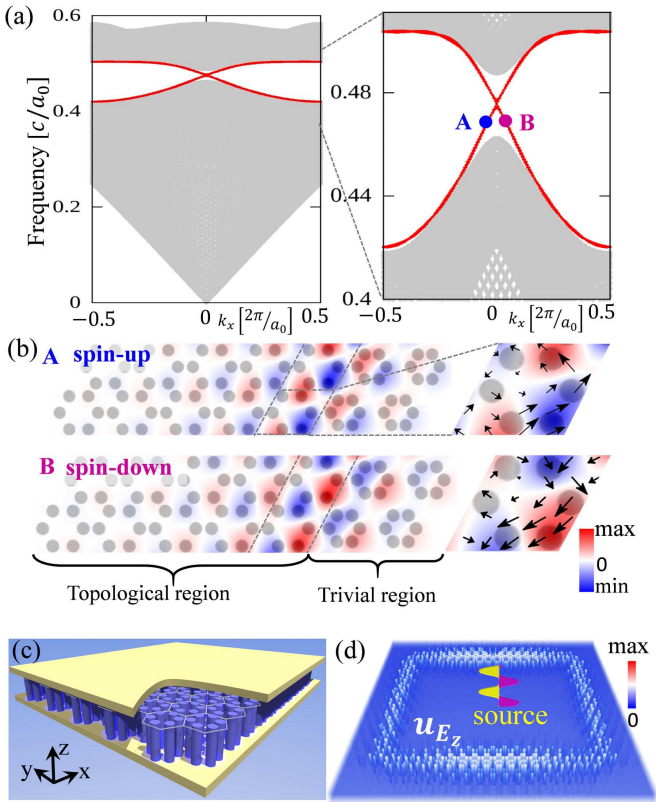}
  \caption{(a) Dispersion relation
  of a ribbon-shaped topological photonic crystal, which is infinite in one
  direction and of 45 and 6 artificial atoms for the topological and trivial
  regions respectively in the other direction. Right panel: enlarged view of
  (a). Red curves are for topological edge states. $A$ and $B$ are two points
  where $E_z$ field are shown in (b). Parameters used here are same as those
  in Fig.~\ref{pdispersion} except $a_0/R=2.9$ and $a_0/R = 3.125$ in
  topological and trivial regions respectively. (b) Real-space distributions of
  $E_z$ fields at points A and B. Right panels: time-averaged Poynting vectors
  $\vec{S}$ over a period. (c) Photonic crystal of height $h$ with two
  horizontal gold plates placed at two ends symmetrically. Hexagons in fade
  white color are primitive cells for the triangular lattice.  (d)
  Distribution of energy-density carried by $E_z$ field $u_{E_z}(\mathbf{r}) =
  \varepsilon(\mathbf{r})|E_z(\mathbf{r})|^2/2$ in a topological
  photonic crystal cladded by a trivial one with $h = 10$ mm, $d = 2.4$ mm and
  $a_0 = 10$ mm, and $R = 3.65$ mm and $3$ mm in topological and trivial regions
  respectively. Frequency of harmonic line source is $\omega = 13.47$ GHz within
  the photonic band gap.}
  \label{ribbon}
\end{figure}

\emph{Topological edge state.---}We also consider a ribbon of photonic crystal
after band inversion by cladding its two edges in terms of two photonic
crystals with trivial band gap (namely before band inversion) at the same
frequency window, which prevents possible edge states from leaking into free
space. It should be kept in mind that, since the cluster of six cylinders is
the basic block of the present design, it should not be destroyed for
any meaningful discussion. As displayed in Fig.~\ref{ribbon}(a), there
appears edge states as indicated by the double degenerated red curves, with a
small gap (not noticeable in this scale) opened at $\Gamma$ point due to the
suppression of $C_6$ symmetry and then pseudo TR symmetry at the interface.
Checking real-space distribution of $E_z$ field at typical momenta around
$\Gamma$ point (A and B in the enlarged vision of Fig.~\ref{ribbon}(a) with
$k_x=\pm 0.04\frac{2\pi}{a_0}$), we find that the in-gap states locate at the
edge and decay exponentially into bulk as displayed in Fig.~\ref{ribbon}(b)
(two other states are localized at the other ribbon edge and not shown
explicitly). As shown in the right insets of Fig.~\ref{ribbon}(b), the
Poynting vectors exhibit a nonzero downward/upward EM energy flow for the
pseudo spin-down/-up state even averaged over time. This indicates
unambiguously counter propagations of EM energy at the sample edge associated
with the two pseudo spin states, the hallmark of a quantum spin Hall effect
(QSHE) state \cite{KaneRMP,SCZhangRMP}. Distributions of Poynting vectors of the bulk bands in
Fig.~\ref{ribbon}(b) for the ribbon system are similar to those
in Fig.~\ref{poynting}(b) for the infinite system. QHE has been described by the
cyclic motions of electrons under strong external magnetic field in a
quasi classic picture of electronic wave functions \cite{CyclicQHE1999}. It is noticed
that the Poynting vector is a physical quantity in EM systems, and
therefore the distributions shown in Figs.~\ref{poynting} and \ref{ribbon}(b)
can be observed in experiments. The photonic QSHE in the present system can also be
confirmed by evaluating the $\mathbb{Z}_2$ topology index based on a $k\cdot
p$ model around $\Gamma$ point. Although Dirac
dispersions in photonic systems were discussed previously in both square and
triangular lattices \cite{CTChan2011,Sakoda2012,SakodaDD}, possible nontrivial
topology was not addressed.

For experimental implementation of the present topological state, the finite
height of silicon rods along $z$ direction has to be taken into account. We
consider explicitly a square sample of topological photonic crystal sandwiched
by two horizontal gold plates separated by $h$ (see Fig.~\ref{ribbon}(c)), with
the height $h$ chosen to prevent photonic bands with nonzero $k_z$ from falling
into the topological band gap. The size of topological sample is
$40\vec{a}_1\times20(\vec{a}_1+\vec{a}_2)$ with all four edges cladded by a
trivial photonic crystal. A harmonic line source $\mathbf{E} = E_0 e^{i\omega
t}\hat{z}$ is placed parallel to silicon rods to inject EM wave at the
interface with the frequency in the topological band gap. We simulate the
system by solving time-dependent Maxwell equations \cite{FDTD,MEEP}. Since any
harmonic source preserves TR symmetry respected by Maxwell equations, the
system exhibits helical topological edge states as shown in
Fig.~\ref{ribbon}(d).

In conclusion, we derive a two-dimensional photonic crystal with nontrivial
topology purely based on silicon, a conventional dielectric material, simply
by deforming honeycomb lattice of silicon cylinders. A pseudo time reversal
symmetry is constructed in terms of the time reversal symmetry respected by
the Maxwell equation in general and the $C_6$ crystal symmetry upon design,
which enables the Kramers doubling with the role of pseudo spin played by the
polarization of magnetic field of transverse magnetic mode. The present
topological photonic crystal with simple design backed up by the symmetry
consideration can be fabricated relatively easy as compared with other
proposals, and is expected to leave impacts to the topological physics and
related materials science.

The authors acknowledge K. Sakoda for useful discussions on photonic Dirac
behaviors. This work was supported by the WPI Initiative on Materials
Nanoarchitectonics, Ministry of Education, Culture, Sports, Science and
Technology of Japan, and partially by Grant-in-Aid for Scientific Research
under the Innovative Area "Topological Quantum Phenomena" (No.25103723),
Ministry of Education, Culture, Sports, Science and Technology of Japan.

\end{document}